\documentclass[prl,superscriptaddress,twocolumn,preprintnumbers,
    showpacs,nofootinbib]{revtex4}

\usepackage{graphicx}% Include figure files
\usepackage{dcolumn}% Align table columns on decimal point
\usepackage{bm}% bold math
\graphicspath{{../Figures-EPS/}}
\usepackage{verbatim}

\begin{document}

\title{Precision study of ground state capture in the $^{14}$N(p,$\gamma$)$^{15}$O reaction}

\author{M.~Marta}\affiliation{Forschungszentrum Dresden-Rossendorf, Bautzner Landstr. 128, 01328 Dresden, Germany}
\author{A.~Formicola}\affiliation{INFN, Laboratori Nazionali del Gran Sasso (LNGS), Assergi (AQ), Italy}
\author{Gy.~Gy\"urky}\affiliation{Institute of Nuclear Research (ATOMKI), Debrecen, Hungary}
\author{D.~Bemmerer}\affiliation{Forschungszentrum Dresden-Rossendorf, Bautzner Landstr. 128, 01328 Dresden, Germany}
\author{C.~Broggini}
 \affiliation{Istituto Nazionale di Fisica Nucleare (INFN), Sezione di Padova, via Marzolo 8, 35131 Padova, Italy}
\author{A.~Caciolli}\affiliation{Istituto Nazionale di Fisica Nucleare (INFN), Sezione di Padova, via Marzolo 8, 35131 Padova, Italy}\affiliation{Dipartimento di Fisica, Universit\`a di Padova, Italy}
\author{P.~Corvisiero}\affiliation{Universit\`a di Genova and INFN Sezione di Genova, Genova, Italy}
\author{H.~Costantini}\affiliation{Universit\`a di Genova and INFN Sezione di Genova, Genova, Italy}
\author{Z.~Elekes}\affiliation{Institute of Nuclear Research (ATOMKI), Debrecen, Hungary}
\author{Zs.~F\"ul\"op}\affiliation{Institute of Nuclear Research (ATOMKI), Debrecen, Hungary}
\author{G.~Gervino}\affiliation{Dipartimento di Fisica Sperimentale, Universit\`a di Torino and INFN Sezione di Torino, Torino, Italy}
\author{A.~Guglielmetti}\affiliation{Istituto di Fisica Generale Applicata, Universit\`a di Milano and INFN Sezione di Milano, Italy}
\author{C.~Gustavino}\affiliation{INFN, Laboratori Nazionali del Gran Sasso (LNGS), Assergi (AQ), Italy}
\author{G.~Imbriani}\affiliation{Dipartimento di Scienze Fisiche, Universit\`a di Napoli ''Federico II'', and INFN Sezione di Napoli, Napoli, Italy}
\author{M.~Junker}\affiliation{INFN, Laboratori Nazionali del Gran Sasso (LNGS), Assergi (AQ), Italy}
\author{R.~Kunz}\affiliation{Institut f\"ur Experimentalphysik III, Ruhr-Universit\"at Bochum, Bochum, Germany}
\author{A.~Lemut}\affiliation{Universit\`a di Genova and INFN Sezione di Genova, Genova, Italy}
\author{B.~Limata}\affiliation{Dipartimento di Scienze Fisiche, Universit\`a di Napoli ''Federico II'', and INFN Sezione di Napoli, Napoli, Italy}
\author{C.~Mazzocchi}\affiliation{Istituto di Fisica Generale Applicata, Universit\`a di Milano and INFN Sezione di Milano, Italy}
\author{R.~Menegazzo}\affiliation{Istituto Nazionale di Fisica Nucleare (INFN), Sezione di Padova, via Marzolo 8, 35131 Padova, Italy}
\author{P.~Prati}\affiliation{Universit\`a di Genova and INFN Sezione di Genova, Genova, Italy}
\author{V.~Roca}\affiliation{Dipartimento di Scienze Fisiche, Universit\`a di Napoli ''Federico II'', and INFN Sezione di Napoli, Napoli, Italy}
\author{C.~Rolfs}\affiliation{Institut f\"ur Experimentalphysik III, Ruhr-Universit\"at Bochum, Bochum, Germany}
\author{M.~Romano}\affiliation{Dipartimento di Scienze Fisiche, Universit\`a di Napoli ''Federico II'', and INFN Sezione di Napoli, Napoli, Italy}
\author{C.~Rossi Alvarez}\affiliation{Istituto Nazionale di Fisica Nucleare (INFN), Sezione di Padova, via Marzolo 8, 35131 Padova, Italy}
\author{E.~Somorjai}\affiliation{Institute of Nuclear Research (ATOMKI), Debrecen, Hungary}
\author{O.~Straniero}\affiliation{Osservatorio Astronomico di Collurania, Teramo, and INFN Sezione di Napoli, Napoli, Italy}
\author{F.~Strieder}\affiliation{Institut f\"ur Experimentalphysik III, Ruhr-Universit\"at Bochum, Bochum, Germany}
\author{F.~Terrasi}\affiliation{Seconda Universit\`a di Napoli, Caserta, and INFN Sezione di Napoli, Napoli, Italy}
\author{H.P.~Trautvetter}\affiliation{Institut f\"ur Experimentalphysik III, Ruhr-Universit\"at Bochum, Bochum, Germany}
\author{A.~Vomiero}\affiliation{INFN, Laboratori Nazionali di Legnaro, Legnaro, Italy}

\collaboration{The LUNA Collaboration}\noaffiliation

\date{\today}
\preprint{as accepted by Phys Rev C rapid communication}

\begin{abstract}
The rate of the hydrogen-burning carbon-nitrogen-oxygen (CNO) cycle is controlled by the slowest process, $^{14}$N(p,$\gamma$)$^{15}$O, which
proceeds by capture to the ground and several excited states in $^{15}$O.
Previous extrapolations for the ground state contribution disagreed by a factor 2, corresponding to 15\% uncertainty in the total astrophysical S-factor.
At the Laboratory for Underground Nuclear Astrophysics (LUNA) 400\,kV accelerator placed deep underground in the Gran Sasso facility in Italy, a new experiment on ground state capture has been carried out at 317.8, 334.4, and 353.3\,keV center-of-mass energy. Systematic corrections have been reduced considerably with respect to previous studies by using a Clover detector and by adopting a relative analysis. The previous discrepancy has been resolved, and ground state capture no longer dominates the uncertainty of the total S-factor.

\end{abstract}


\maketitle

Recent data on the abundance of the elements carbon, nitrogen, and oxygen (CNO) in the solar atmosphere \cite{Asplund06-NPA} lead to a contradiction between solar model predictions and measurements for several helioseismological quantities \cite{Bahcall06-ApJS}.
In the present precision era, this puzzle represents the foremost problem of the standard solar model \cite{Bahcall06-ApJS} since the resolution of the solar neutrino puzzle \cite{Ahmad02-PRL}. In order to address this point, it has been suggested to determine the CNO abundances in the solar center from neutrino data \cite{Haxton07-arxiv}.
Neutrinos emitted in solar CNO cycle burning are expected to lead to about 1000 events/year both in the Borexino detector \cite{Arpesella07-PLB} and in the proposed SNO+ detector \cite{Chen05-NPBPS}. A correct interpretation of this expected data, based on the known solar core temperature and known neutrino properties \cite{Haxton07-arxiv}, requires the rate of the CNO cycle to be known with systematical uncertainty matching these statistics.

\begin{figure}[tb]
\centering
 \includegraphics[angle=00,width=1.0\columnwidth]{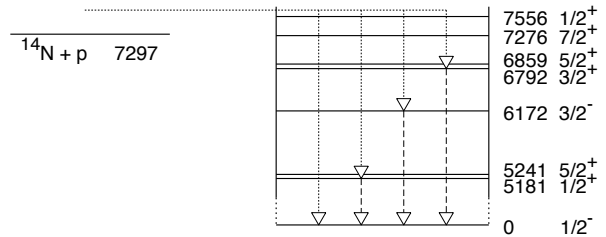}
 \caption{\label{fig:Levels-O15} Energy levels of $^{15}$O, in keV \cite{Ajzenberg91-NPA,Imbriani05-EPJA}.}
\end{figure}

The rate of the CNO cycle is controlled \cite{Iliadis07}
by the $^{14}$N(p,$\gamma$)$^{15}$O reaction.
Its cross section $\sigma(E)$, parameterized%
\footnote{$E_{\rm p}$ denotes the beam energy in the laboratory system, and $E$ the effective energy in the center of mass system in keV.}
as the astrophysical S-factor
\begin{equation}
S(E) = \sigma E \exp \left[212.4/\sqrt{E} \right],
\end{equation}
has been extensively studied in the past \cite[and references therein]{Schroeder87-NPA}. Recently, it has been shown that capture to the ground state in $^{15}$O (fig.~\ref{fig:Levels-O15}), previously \cite{Schroeder87-NPA} believed to account for half of the S-factor% 
\footnote{$S_i(0)$ denotes the S-factor, extrapolated to zero energy, for capture to the state at $i$~keV in $^{15}$O. $S_{\rm GS}(0)$ and $S_{\rm tot}(0)$ refer to ground state capture and to the total S-factor, respectively.}
extrapolated to zero energy $S_{\rm tot}(0)$, is strongly suppressed \cite{Bertone01-PRL,Angulo01-NPA,Mukhamedzhanov03-PRC,Yamada04-PLB,Formicola04-PLB,Runkle05-PRL,Imbriani05-EPJA}. This finding is independently supported by a reduction in the $\gamma$-width of the subthreshold state at 6792\,keV in $^{15}$O seen in Doppler shift attenuation \cite{Bertone01-PRL} and Coulomb excitation \cite{Yamada04-PLB} works, and by fits \cite{Angulo01-NPA,Mukhamedzhanov03-PRC,Formicola04-PLB,Runkle05-PRL,Imbriani05-EPJA} in the R-matrix framework (table~\ref{tab:Gammawidths}). The resulting 50\% reduction in the total cross section has subsequently been directly observed at an energy as low as $E$ $\approx$ 70\,keV \cite{Lemut06-PLB}.

For the Gamow peak of the Sun ($E$ $\approx$ 27\,keV), however, extrapolations remain indispensable. For the dominant contribution to $S_{\rm tot}(0)$, i.e. capture to the state at 6792\,keV, recent experimental data and R-matrix fits are consistent \cite{Runkle05-PRL,Imbriani05-EPJA}.
For capture to the ground state, recent experimental data ($E$ $\approx$ 120-480\,keV) from LUNA \cite{Formicola04-PLB,Imbriani05-EPJA} and TUNL \cite{Runkle05-PRL} are consistent with each other, and they both rule out a previous R-matrix fit \cite{Mukhamedzhanov03-PRC}. However, the extrapolated $S_{\rm GS}(0)$ values \cite{Formicola04-PLB,Runkle05-PRL} disagree significantly (table~\ref{tab:Gammawidths}). This discrepancy has 15\% impact on $S_{\rm tot}(0)$, limiting its precision.
In addition to differently treating previous data \cite{Schroeder87-NPA} in the fit,
Refs.~\cite{Formicola04-PLB,Runkle05-PRL} had employed large germanium detectors in close geometry, enhancing the detection efficiency but incurring true coincidence summing-in corrections of 100-250\% for the ground state data, which, in turn, lead to considerable systematic uncertainty.

% ==================
\begin{table}[tb]
\centering
\caption{Measured quantities used to obtain an extrapolated $S_{\rm GS}(0)$ [keV barn] in recent studies.}
\label{tab:Gammawidths}
\begin{ruledtabular}
\begin{tabular}{llc}
Group & Quantity used [taken from] & $S_{\rm GS}(0)$\\ \hline
TUNL \cite{Bertone01-PRL} & $\gamma$-width \cite{Bertone01-PRL} & 0.12--0.45 \\
Brussels \cite{Angulo01-NPA} & Cross section \cite{Schroeder87-NPA} & $0.08 \! \begin{array}{c} \scriptstyle + 0.13 \\[-1mm] \scriptstyle - 0.06 \end{array}$ \\
Texas A\&M \cite{Mukhamedzhanov03-PRC} & ANC \cite{Mukhamedzhanov03-PRC}, cross section \cite{Schroeder87-NPA} & 0.15$\pm$0.07 \\
LUNA \cite{Formicola04-PLB} & Cross section \cite{Schroeder87-NPA,Formicola04-PLB}\footnote{Ref.~\cite{Schroeder87-NPA} data have been corrected \cite{Formicola04-PLB} for summing-in.} & 0.25$\pm$0.06 \\
TUNL \cite{Runkle05-PRL} & Cross section \cite{Runkle05-PRL} & 0.49$\pm$0.08 \\
\end{tabular}
\end{ruledtabular}
\end{table}
%
% ==================

The aim of the present work is to address the conflicting extrapolations \cite{Formicola04-PLB,Runkle05-PRL} with a precision cross section measurement. In order to minimize the uncertainties, the analysis is limited to the ratio of the cross sections for capture to the ground state and to the 6792\,keV state. An energy range above the 259\,keV resonance, where the fits for ground state capture pass through a sensitive minimum \cite{Angulo01-NPA}, has been selected \cite{Trautvetter08-NPA3}. 
A second sensitive energy region lies below the 259\,keV resonance. Since the cross section is a factor 100 lower there, the latter energies were not probed in the present work.
The experiment was performed at the Laboratory for Underground Nuclear Astrophysics (LUNA) at the Gran Sasso National Laboratory (Italy), which has ultra-low $\gamma$-ray laboratory background \cite{Bemmerer05-EPJA}.
A Clover detector was used, reducing the summing-in correction by a factor 30 (table~\ref{tab:Results}).

The H$^+$ beam of $E_{\rm p}$ = 359, 380, and 399\,keV and 0.25-0.45\,mA intensity from the LUNA2 400\,kV accelerator \cite{Formicola03-NIMA} impinged on a sputtered TiN target, with 55\,keV thickness measured on the $E$ = 259\,keV resonance. The $\gamma$-rays from the reaction to be studied were detected in a Eurisys Clover-BGO detection system \cite{Elekes03-NIMA}. The front end of the Clover crystals was positioned at 9.5\,cm distance from the target, at an angle of 55$^\circ$ with respect to the beam axis. The output signal from each of the four Clover segments was split into two branches; of these branches, one branch was recorded separately, and the four spectra were summed in the offline analysis (singles mode). The second branches of the four signals were added online in an analog summing unit (addback mode). For experiments off the 259\,keV resonance, the addback mode data were recorded in anticoincidence with the BGO anti-Compton shield.

%%%%%%%%%%%%%%%%%%%%%%%%%%%%%%%%
%%%%%%%%%%%%%%%%%%%%%%%%%%%%%%%%

The $\gamma$-ray detection efficiency was obtained using $^{137}$Cs and $^{60}$Co radioactive sources calibrated to 1.5\% and 0.75\%, respectively. The efficiency curve was extended to high energy based on spectra recorded at the 259\,keV resonance, using the known 1:1 $\gamma$-ray cascades for the excited states at 6172 and 6792\,keV. The $\gamma$-rays from the decay of this 1/2$^+$ resonance are isotropic, and their angular correlations are well known \cite{PovhHebbard59-PR}. The calculated summing-out correction in addback mode is 2.9\%, with an assumed relative uncertainty of 20\%, consistent with a GEANT4 \cite{Agostinelli03-NIMA} simulation showing (4.5$\pm$1.8)\% correction. As a check on the quality of the efficiency curve, the experimental cascade ratio for the 5181\,keV excited state (not used in the fit) was found to be reproduced within 1\% statistics.

%%%%%%%%%%%%%%%%%%%%%%%%%%%%%%%%
\begin{figure*}[t!!]
\centering
 \includegraphics[angle=270,width=1.0\textwidth]{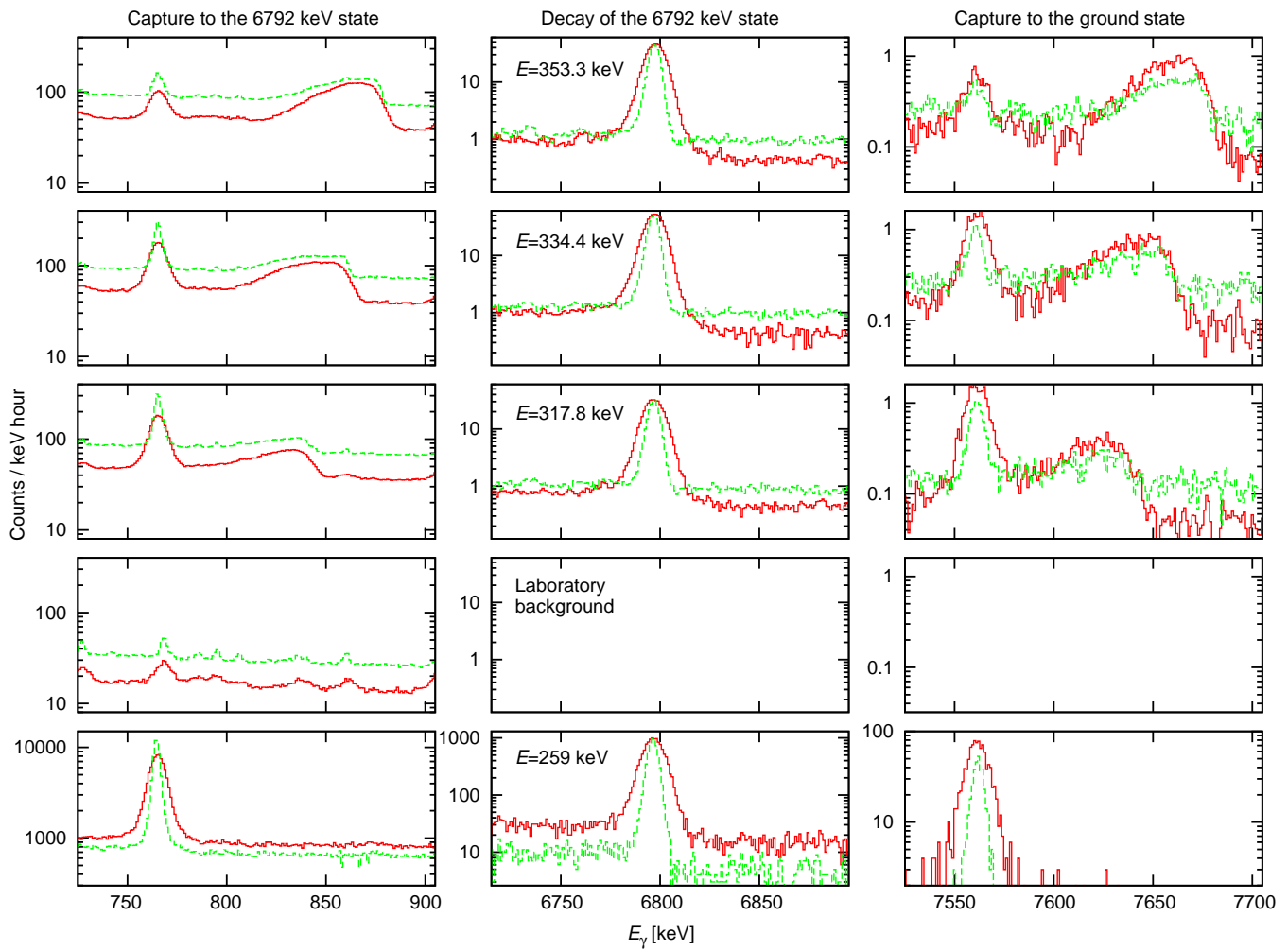}
 \caption{\label{fig:Cloverspec} (color online) Solid red (dashed green) line: $\gamma$-ray spectra for addback (singles) mode. First three rows: off-resonance data. Fourth row: laboratory background, negligible at high $\gamma$-energy. Fifth row: data at the $E$ = 259\,keV resonance.}
\end{figure*}
%%%%%%%%%%%%%%%%%%%%%%%%%%%%%%%%

The branching ratio for decay of the 259\,keV resonance to the ground state was found to be (1.56$\pm$0.08)\% in addback mode and (1.53$\pm$0.06)\% in singles mode, taking into account (42$\pm$2)\% and (7.4$\pm$0.3)\% summing-in correction, respectively. This confirms that the summing-in correction for the addback mode is accurate. 
Furthermore, the GEANT4 simulation showed (40.2$\pm$1.4)\% and (7.8$\pm$0.9)\% summing-in correction for addback and singles, respectively, in good agreement with the above data.
The branching ratio is in good agreement with the previous LUNA value \cite{Imbriani05-EPJA} and in fair agreement with TUNL \cite{Runkle05-PRL}.

Off resonance, the spectra (fig.~\ref{fig:Cloverspec}, rows 1-3) show some on-resonance contribution due to the tail of the target profile. The secondary $\gamma$-ray from the decay of the 6792\,keV level (fig.~\ref{fig:Cloverspec}, middle column) therefore contains 13-55\% on-resonance capture, and it was rescaled with the on/off-resonance ratio obtained from the primary $\gamma$-rays (fig.~\ref{fig:Cloverspec}, left column). Subsequently, the cross section ratio
\begin{equation}
R_{\rm GS/6792}(E) = \frac{\sigma_{\rm GS}(E)}{\sigma_{\rm 6792}(E)}
\end{equation}
with $\sigma_{\rm GS}(E)$ and $\sigma_{\rm 6792}(E)$ the cross sections for capture to the ground state and to the 6792\,keV state in $^{15}$O, respectively, was calculated for each bombarding energy (table~\ref{tab:Results}).
The addback and singles mode data for $R_{\rm GS/6792}$ were found to be in agreement. Because of their lower statistical uncertainty, the addback data were adopted for the further analysis.

The systematic uncertainty for $R_{\rm GS/6792}$ (table~\ref{tab:Results}) depends on (1) the summing-in correction for the ground state $\gamma$-ray (up to 4.6\% and 0.9\% effect on $R_{\rm GS/6792}$ for the addback and singles mode, respectively, taking into account the calculated \cite{Iliadis07} angular correlation), and (2) the slope of the detection efficiency curve over the energy range $E_\gamma$ = 6792-7650\,keV (known to 0.8\%). For the cascade 6792\,keV $\gamma$-ray, (3) the anticoincidence efficiency (1.2\% effect), and (4) the summing-out correction (0.6\% effect) contribute to the systematic uncertainty for $R_{\rm GS/6792}$. The effects of e.g. target composition and profile, stopping power, beam intensity, and absolute $\gamma$-ray detection efficiency cancel out in the relative experiment. The effective energy $E$ was determined from the centroids of the $\gamma$-lines for capture to the ground state and to the 6792\,keV state and leads to 2.4\% uncertainty.

%%%%%%%%%%%%%%%%%%%%%%%%%%%%%%%%
 \begin{table}[tb]
\caption{\label{tab:Results} Cross section ratio $R_{\rm GS/6792}(E)$ and relative uncertainty. The size of the summing-in correction is also given.}
\begin{ruledtabular}
\begin{tabular}{clcccc}
        $E$ [keV] & mode & $R_{\rm GS/6792}(E)$ &
        stat. & syst. & Summing-in \\
& & [10$^{-2}$] & \multicolumn{2}{c}{uncertainty}& correction \\
        \hline
317.8$\pm$1.5 & addback & 4.71 & 5.9\% & 5.4\% & 30\% \\
& singles & 4.67 & 14\% & 2.7\% & 4.3\% \\
334.4$\pm$1.5 & addback & 5.00 & 5.1\% & 3.9\% & 21\% \\
& singles & 5.07 & 13\% & 2.5\% & 3.4\% \\
353.3$\pm$1.5 & addback & 5.30 & 3.6\% & 3.5\% & 19\% \\
& singles & 5.15 & 10\% & 2.3\% & 3.2\% \\
\end{tabular}
\end{ruledtabular}
\end{table}
%%%%%%%%%%%%%%%%%%%%%%%%%%%%%%%%

%%%%%%%%%%%%%%%%%%%%%%%%%%%%%%%%
\begin{figure}[t!!]
\centering
 \includegraphics[angle=-90,width=1.0\columnwidth]{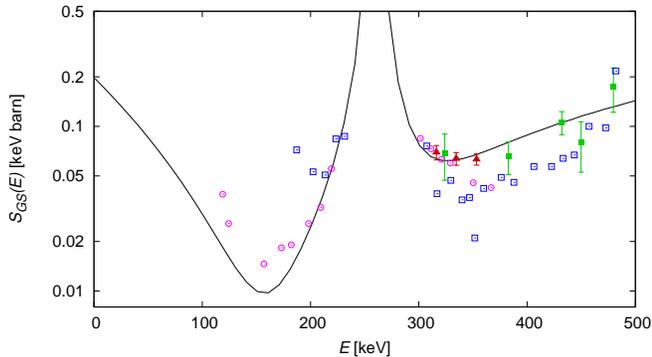}
 \caption{\label{fig:Rmatrix} 
 S-factor for capture to the ground state. Full triangles: present data. Full squares: Ref.~\cite{Schroeder87-NPA}. Line: Present best R-matrix fit. Data from \cite{Formicola04-PLB,Imbriani05-EPJA} (empty circles) and \cite{Runkle05-PRL} (empty squares) are shown for comparison but were not used in the fit; their error bars have been omitted for clarity.
 }
\end{figure}

%%%%%%%%%%%%%%%%%%%%%%%%%%%%%%%%

The absolute cross section for the ground state transition obtained from the present data was determined by the ratios given in table~\ref{tab:Results} normalized with the weighted average (uncertainty 7.5\%) of the S-factor results for the 6792~keV transition given in Refs.~\cite{Schroeder87-NPA,Runkle05-PRL,Imbriani05-EPJA}. 
From such a combined fit an ANC of 4.8\,fm$^{-1/2}$ was obtained for the 6792~keV state, in good agreement with Refs.~\cite{Mukhamedzhanov03-PRC, Bertone02-PRC} and resulting in
$\gamma^2$ = 0.4~MeV for the reduced width of the subthreshold state.
For the strength of the 259\,keV resonance, 13.1\,meV (weighted average of \cite{Runkle05-PRL,Imbriani05-EPJA,Lemut06-PLB,Ajzenberg91-NPA}) was adopted, 
for its proton width 0.99\,keV \cite{Formicola04-PLB},
and for the ground state branching, 1.63\% (weighted average of \cite{Runkle05-PRL,Imbriani05-EPJA} and the present work) was used. 
For all other parameters, the previous values were taken without any change \cite{Formicola04-PLB}: 
ANC for ground state capture: 7.3\,fm$^{-1/2}$.
$E$ = 0.987\,MeV resonance: $\Gamma_\gamma$ = 26\,meV, $\Gamma_{\rm p}$ = 3\,keV. 
$E$ = 2.187\,MeV resonance: $\Gamma_\gamma$ = 4.4\,eV, $\Gamma_{\rm p}$ = 0.27\,MeV. 
Background pole at $E$ = 6\,MeV, $\Gamma_{\rm p}$ = 8\,MeV. 
In order to limit the systematic uncertainty due to summing-in to less than the statistical error, only data with less than 50\% summing-in correction were used for the R-matrix analysis: i.e. \cite{Schroeder87-NPA} (corrected \cite{Formicola04-PLB} for summing-in) and the present data. The interference pattern around the 259\,keV resonance is fixed by the results of \cite{Formicola04-PLB,Runkle05-PRL,Imbriani05-EPJA}, and the interaction radius was set to 5.5~fm \cite{Formicola04-PLB}. 
The best fit (fig.~\ref{fig:Rmatrix}) varying only the $\gamma$-widths of the subthreshold state and of the background pole results in $S_{\rm GS}(0)$ = 0.20\,keV\,barn with a $\gamma$-width $\Gamma_{\gamma}$ = 0.9$\pm$0.2\,eV for the subthreshold state, in agreement with Coulomb excitation work \cite{Yamada04-PLB} and with lifetime measurements \cite{Bertone01-PRL,Schuermann08-PRC}. A full R-matrix analysis including a detailed error determination for all parameters is beyond the scope of the present work. Therefore, the previous relative uncertainty of 24\% in $S_{\rm GS}(0)$ \cite{Formicola04-PLB} is adopted here, giving $S_{\rm GS}(0)$ = 0.20$\pm$0.05\,keV\,barn. 

% In order to check the robustness of the $S_{\rm GS}(0)$ determination, the previous TUNL \cite{Runkle05-PRL} R-matrix analysis was repeated, based \cite{Runkle05-PRL} on the TUNL data and the present relative data, rescaled with the TUNL R-matrix fit for capture to the 6792 keV state. All parameters for the R-matrix calculation were taken from Ref. \cite{Runkle05-PRL} except for the $\gamma$-widths of the subthreshold state and the background pole which were varied. Moreover, a ground state ANC of 7.5\,fm$^{-1/2}$ (weighted mean of \cite{Mukhamedzhanov03-PRC,Bertone02-PRC}) has been adopted, because no explicit ANC has been given in Ref. \cite{Runkle05-PRL}. The extrapolation is then $S_{\rm GS}(0)$ = 0.23~keV\,barn, without including the Ref.~\cite{Schroeder87-NPA} data in the fit. Hence, the high extrapolated value $S_{\rm GS}(0)$ = 0.49$\pm$0.08~keV\,barn \cite{Runkle05-PRL} is ruled out. Concluding, the best R-matrix fit (fig.~\ref{fig:Rmatrix}) results in $S_{\rm GS}(0)$ = 0.20$\pm$0.05~keV\,barn, in good agreement with the previous LUNA value \cite{Formicola04-PLB,Imbriani05-EPJA}.

In summary, owing to the present high precision data, ground state capture now contributes less than 4\% uncertainty to the total $S_{\rm tot}(0)$, instead of the previous 15\%, based on a data set which is nearly free from summing problems. On the basis of the present result, $S_{\rm tot}(0)$ = 1.57$\pm$0.13\,keV\,barn is recommended, with the uncertainty including also systematic effects.  For this sum, $S_{\rm 6172}$(0) = 0.09$\pm$0.07\,keV\,barn \cite{Nelson03-PRC,Mukhamedzhanov03-PRC,Runkle05-PRL,Imbriani05-EPJA} has been adopted. Further improvements in $S_{\rm tot}(0)$ precision would require a fresh study of this contribution.
In the meantime, the present ground state data pave the way for a measurement of the solar central metallicity \cite{Haxton07-arxiv}.

\begin{acknowledgments}
The use of the R-matrix code \cite{Angulo01-NPA} written by P. Descouvemont (ULB Brussels) is gratefully acknowledged. 
One of us (H.P.T.) thanks R.E. Azuma, E. Simpson, and A. Champagne for fruitful discussions. 
--- The present work has been supported by INFN and in part by the EU (ILIAS-TA RII3-CT-2004-506222), OTKA (T49245 and K68801), and DFG (Ro~429/41).
\end{acknowledgments}

%%%%%%%%%%%%%%%%%%%%%%%%%%%%%%%%

\end{document}